\documentclass[12pt,twoside]{article}

\usepackage{amsmath}

\usepackage{amssymb}

\date{}

\begin{document}

\centerline{}

\centerline{}

\centerline {\Large{\bf A Description of the Subgraph Induced at }}

\centerline{}

\centerline{\Large{\bf a Labeling of a Graph by the Subset of }}

\centerline{}

\centerline{\Large{\bf Vertices with an Interval Spectrum.}}

\centerline{}

\centerline{\bf {Narine N. Davtyan}}

\centerline{}

\centerline{Ijevan Branch of Yerevan State University, Ijevan,
Republic of Armenia}

\centerline{nndavtyan@gmail.com}

\centerline{}

\centerline{\bf {Arpine M. Khachatryan}}

\centerline{}

\centerline{Ijevan Branch of Yerevan State University, Ijevan,
Republic of Armenia}

\centerline{khachatryanarpine@gmail.com}

\centerline{}

\centerline{\bf {Rafayel R. Kamalian}}

\centerline{}

\centerline{Ijevan Branch of Yerevan State University, Ijevan}

\centerline{The Institute for Informatics and Automation Problems of
NAS RA,}

\centerline{Yerevan, Republic of Armenia, rrkamalian@yahoo.com}

\newtheorem{Theorem}{\quad Theorem}[section]

\newtheorem{Definition}[Theorem]{\quad Definition}

\newtheorem{Corollary}[Theorem]{\quad Corollary}

\newtheorem{Lemma}[Theorem]{\quad Lemma}

\newtheorem{Example}[Theorem]{\quad Example}

\centerline{}

\begin{abstract}
The sets of vertices and edges of an undirected, simple, finite,
connected graph $G$ are denoted by $V(G)$ and $E(G)$, respectively.
An arbitrary nonempty finite subset of consecutive integers is
called an interval. An injective mapping $\varphi:E(G)\rightarrow
\{1,2,\dots,|E(G)|\}$ is called a labeling of the graph $G$. If $G$
is a graph, $x$ is its arbitrary vertex, and $\varphi$ is its
arbitrary labeling, then the set $S_G(x,\varphi)\equiv\{\varphi(e)/
e\in E(G), e \textrm{ is incident with } x$\} is called a spectrum
of the vertex $x$ of the graph $G$ at its labeling $\varphi$. For
any graph $G$ and its arbitrary labeling $\varphi$, a structure of
the subgraph of $G$, induced by the subset of vertices of $G$ with
an interval spectrum, is described.
\end{abstract}



\section{Introduction}

We consider undirected, simple, finite and connected graphs,
containing at least one edge. The terms and concepts which are not
defined can be found in \cite{West}.

For a graph $G$, we denote by $V(G)$ and $E(G)$ the sets of its
vertices and edges, respectively. The set of vertices of $G$
adjacent to a vertex $x\in V(G)$ is denoted by $I_G(x)$. The set of
edges of $G$ incident with a vertex $x\in V(G)$ is denoted by
$J_G(x)$.

If $G$ is a graph, and $x\in V(G)$, then $d_G(x)$ denotes the degree
of the vertex $x$ in the graph $G$. For any graph $G$, $\Delta(G)$
denotes the maximum degree of a vertex of $G$.

For any graph $G$, we define the subsets $V'(G)$ and $V''(G)$ of its
vertices by the following way: $V'(G)\equiv\{x\in
V(G)/\;d_G(x)=1\}$, $V''(G)\equiv\{x\in V(G)/\;d_G(x)\geq2\}$.

The distance in a graph $G$ between its vertices $x$ and $y$ is
denoted by $d_G(x,y)$.

For any graph $G$, we denote by $diam(G)$ its diameter. A vertex
$x\in V(G)$ is called a peripheral vertex of a graph $G$ if there
exists a vertex $y\in V(G)$ satisfying the condition
$d_G(x,y)=diam(G)$.

For an arbitrary nonempty finite subset $A$ of the set
$\mathbb{Z}_+$, we denote by $l(A)$ and $L(A)$, respectively, the
least and the greatest element of $A$.

An arbitrary nonempty finite subset $A$ of the set $\mathbb{Z}_+$ is
called an interval if it satisfies the condition $|A|=L(A)-l(A)+1$.
An interval with the minimum element $p$ and the maximum element $q$
is denoted by $[p,q]$.

An injective mapping $\varphi:E(G)\rightarrow \mathbb{N}$ is called
a labeling of the graph $G$. For any graph $G$, we denote by
$\psi(G)$ the set of all labelings of the graph $G$.

If $G$ is a graph, $\varphi\in\psi(G)$, and $E_0\subseteq E(G)$, we
set
$$
\varphi[E_0]\equiv\bigcup_{e\in E_0}\{\varphi(e)\}.
$$

If $G$ is a graph, $x\in V(G)$, and $\varphi\in\psi(G)$, then the
set $\varphi[J_G(x)]$ is called a spectrum of the vertex $x$ of the
graph $G$ at the labeling $\varphi$.

If $G$ is a graph, and $\varphi\in\psi(G)$, then we set
$U(G,\varphi)\equiv\{x\in V(G)/\;\varphi[J_G(x)] \\ \textrm{ is an
interval}\}$.

If $G$ is a graph, and $\varphi\in\psi(G)$, then we denote by
$G^{(\varphi, int)}$ the subgraph of the graph $G$ induced by the
subset $U(G,\varphi)$ of its vertices.

For any graph $G$, we set
$\lambda(G)\equiv\{\varphi\in\psi(G)/\;U(G,\varphi)\neq\emptyset\}$.
Clearly, for any graph $G$, $\lambda(G)\neq\emptyset$.

If $G$ is a graph, $\varphi\in\lambda(G)$, $(x_0,x_1)\in E(G)$, then
the simple path $P=(x_0,(x_0,x_1),x_1)$ is called a trivial
$\varphi$-gradient path of $G$ iff the following two conditions
hold:
\begin{enumerate}
  \item $\{x_0,x_1\}\subseteq U(G,\varphi)$,
  \item at least one of the following two conditions holds:
  \begin{enumerate}
    \item
    $\varphi((x_0,x_1))=L(\varphi[J_{G^{(\varphi, int)}}(x_0)])=
        l(\varphi[J_{G^{(\varphi, int)}}(x_1)])$,
    \item
    $\varphi((x_0,x_1))=l(\varphi[J_{G^{(\varphi, int)}}(x_0)])=
        L(\varphi[J_{G^{(\varphi, int)}}(x_1)])$.
  \end{enumerate}
\end{enumerate}

If $G$ is a graph, $\varphi\in\lambda(G)$, then a simple path
$P=(x_0,(x_0,x_1),x_1,\dots,x_k,\\(x_k,x_{k+1}),x_{k+1})$ with $k\in
\mathbb{Z}_+$ is called a $\varphi$-gradient path of $G$, if either
$k=0$ and $P$ is a trivial $\varphi$-gradient path of $G$, or $k\in
\mathbb{N}$ and the following two conditions hold:
\begin{enumerate}
    \item $V(P)\subseteq U(G,\varphi)$,
    \item exactly one of the following two conditions holds:
    \begin{enumerate}
    \item for any $i\in[0,k]$,
    $\varphi((x_i,x_{i+1}))=L(\varphi[J_{G^{(\varphi, int)}}(x_i)])=
        l(\varphi[J_{G^{(\varphi, int)}}(x_{i+1})])$,
    \item for any $i\in[0,k]$,
    $\varphi((x_i,x_{i+1}))=l(\varphi[J_{G^{(\varphi, int)}}(x_i)])=
        L(\varphi[J_{G^{(\varphi, int)}}(x_{i+1})])$.
    \end{enumerate}
\end{enumerate}

If $G$ is a graph, $\varphi\in\lambda(G)$, then the set of all
$\varphi$-gradient paths of $G$ is denoted by $\xi(G,\varphi)$.

If $G$ is a graph, $\varphi\in\lambda(G)$, and $P\in\xi(G,\varphi)$,
then $P$ is called a maximal $\varphi$-gradient path of $G$, if
there is no $\widetilde{P}\in\xi(G,\varphi)$ with $V(P)\subset
V(\widetilde{P})$.

If $G$ is a graph, $\varphi\in\lambda(G)$, then the set of all
maximal $\varphi$-gradient paths of $G$ is denoted by
$\tau(G,\varphi)$.

For arbitrary integers $n$ and $i$, satisfying the inequalities
$n\geq3$, $2\leq i\leq n-1$, and for any sequence
$A_{n-2}\equiv(a_1,a_2,\dots,a_{n-2})$ of nonnegative integers, we
define the sets $V[i,A_{n-2}]$ and $E[i,A_{n-2}]$ as follows:
$$
\begin{array}{l}
V[i,A_{n-2}]\equiv\left\{
\begin{array}{ll}
\{y_{i,1},\dots,y_{i,a_{i-1}}\}, & \textrm{if $\;a_{i-1}>0$}\\
\emptyset, & \textrm{if $\;a_{i-1}=0$,}
\end{array}
\right.
\\
E[i,A_{n-2}]\equiv\left\{
\begin{array}{ll}
\{(x_i,y_{i,j}),/\;1\leq j\leq a_{i-1}\}, & \textrm{if $\;a_{i-1}>0$}\\
\emptyset, & \textrm{if $\;a_{i-1}=0$.}
\end{array}
\right.
\end{array}
$$

For any integer $n\geq3$, and for any sequence
$A_{n-2}\equiv(a_1,a_2,\dots,a_{n-2})$ of nonnegative integers, we
define a graph $T[A_{n-2}]$ as follows:
$$
\begin{array}{l}
V(T[A_{n-2}])\equiv\{x_1,\dots,x_n\}\cup\Big(\bigcup_{i=2}^{n-1}V[i,A_{n-2}]\Big),\\
E(T[A_{n-2}])\equiv\{(x_i,x_{i+1})/\;1\leq i\leq
n-1\}\cup\Big(\bigcup_{i=2}^{n-1}E[i,A_{n-2}]\Big).
\end{array}
$$

A graph $G$ is called a galaxy, if either $G\cong K_2$, or there
exist an integer $n\geq3$ and a sequence
$A_{n-2}\equiv(a_1,a_2,\dots,a_{n-2})$ of nonnegative integers, for
which $G\cong T[A_{n-2}]$.

In the paper, for any graph $G$ and arbitrary
$\varphi\in\lambda(G)$, a structure of the subgraph
$G^{(\varphi,int)}$ of the graph $G$ is described. The main result
was announced in \cite{Luhansk}.

\section{Preliminary Notes}

\begin{Lemma}\label{l1}
If $G$ is a graph, $\varphi\in\lambda(G)$, $\{x,y\}\subseteq
U(G,\varphi)$, $(x,y)\in E(G)$, then $|\varphi[J_G(x)]\cap
\varphi[J_G(y)]|=1$.
\end{Lemma}

{\it Proof.} If $\min\{d_G(x),d_G(y)\}=1$, the statement is evident.
Now suppose that $\min\{d_G(x),d_G(y)\}\geq2$. Since $(x,y)\in E(G)$
we have $|\varphi[J_G(x)]\cap \varphi[J_G(y)]|\geq1$. Let us assume
that $|\varphi[J_G(x)]\cap \varphi[J_G(y)]|\geq2$. It means that
there exist $e'\in J_G(x)$, $e''\in J_G(y)$, which satisfy the
conditions $e'\neq(x,y)$, $e''\neq(x,y)$, $e'\neq e''$,
$\varphi(e')=\varphi(e'')$. It is incompatible with
$\varphi\in\lambda(G)$.

{\it Lemma is proved.}

\begin{Corollary}\label{c1}
Let $G$ be a graph, and $\varphi\in\lambda(G)$. Suppose that
vertices $x_1,x_2,\dots,x_n$ $(n\geq2)$ of the graph $G$ satisfy the
conditions
\begin{enumerate}
  \item $\{x_1,\dots,x_n\}\subseteq U(G,\varphi)\cap V''(G)$,
  \item for any $i\in[1,n-1]$, $(x_i,x_{i+1})\in E(G)$.
\end{enumerate}
Then exactly one of the following two statements is true:
\begin{enumerate}
  \item for any $i\in[1,n-1]$, $\varphi((x_i,x_{i+1}))=L(\varphi[J_G(x_i)])=l(\varphi[J_G(x_{i+1})])$,
  \item for any $i\in[1,n-1]$, $\varphi((x_i,x_{i+1}))=l(\varphi[J_G(x_i)])=L(\varphi[J_G(x_{i+1})])$.
\end{enumerate}
\end{Corollary}

\begin{Corollary}\label{c2}
If $G$ is a graph, and $\varphi\in\lambda(G)$, then
$G^{(\varphi,int)}$ is a forest.
\end{Corollary}

{\it Proof.} Assume the contrary: the graph $G^{(\varphi,int)}$
contains a subgraph $G_0$ which is isomorphic to a simple cycle.
Clearly, there exists an edge $e_0=(x,y)\in E(G_0)$, for which
$\varphi(e_0)=l(\varphi[E(G_0)])$. From here, taking into account
that $\{x,y\}\subseteq U(G,\varphi)\cap V''(G)$, we obtain
$|\varphi[J_G(x)]\cap \varphi[J_G(y)]|\geq2$. It contradicts lemma
\ref{l1}.

{\it Corollary is proved.}

\begin{Lemma}\label{l2}
Let $G$ be a graph, and $\varphi\in\lambda(G)$. Then, for an
arbitrary vertex $x\in U(G,\varphi)$, the inequality $|I_G(x)\cap
U(G,\varphi)\cap V''(G)|\leq2$ holds.
\end{Lemma}

{\it Proof.} Suppose that there exists a vertex $z_0\in
U(G,\varphi)$ with $|I_G(z_0)\cap U(G,\varphi)\cap V''(G)|\geq3$.
Let us choose three different vertices $y_1,y_2,y_3$ from the set
$I_G(z_0)\cap U(G,\varphi)\cap V''(G)$.

Note that the vertices $y_1,z_0,y_2$ of the graph $G$ satisfy the
conditions of corollary \ref{c1} (with $y_1$ in the role of $x_1$,
$z_0$ in the role of $x_2$, $y_2$ in the role of $x_3$, and with
$n=3$).

Note also that the vertices $y_1,z_0,y_3$ of the graph $G$ satisfy
the conditions of corollary \ref{c1} (with $y_1$ in the role of
$x_1$, $z_0$ in the role of $x_2$, $y_3$ in the role of $x_3$, and
with $n=3$).

{\it Case 1.} For the vertices $y_1,z_0,y_2$ of the graph $G$, the
statement 1) of corollary \ref{c1} is true. It means that
$\varphi((y_1,z_0))=L(\varphi[J_G(y_1)])=l(\varphi[J_G(z_0)])$,
$\varphi((z_0,y_2))=L(\varphi[J_G(z_0)])=l(\varphi[J_G(y_2)])$. It
is not difficult to see that for the vertices $y_1,z_0,y_3$ of the
graph $G$ also the statement 1) of corollary \ref{c1} is true. It
means that
$\varphi((y_1,z_0))=L(\varphi[J_G(y_1)])=l(\varphi[J_G(z_0)])$,
$\varphi((z_0,y_3))=L(\varphi[J_G(z_0)])=l(\varphi[J_G(y_3)])$. The
equalities $\varphi((z_0,y_2))=L(\varphi[J_G(z_0)])$ and
$\varphi((z_0,y_3))=L(\varphi[J_G(z_0)])$ are incompatible.

{\it Case 2.} For the vertices $y_1,z_0,y_2$ of the graph $G$, the
statement 2) of corollary \ref{c1} is true.

The proof is similar as in case 1.

{\it Lemma is proved.}

\begin{Lemma}\label{l3}
Let $G$ be a graph, $\varphi\in\lambda(G)$,
$\xi(G,\varphi)\neq\emptyset$, $P\in\xi(G,\varphi)$. Then there
exists a unique $\widetilde{P}\in\tau(G,\varphi)$ satisfying the
condition $V(P)\subseteq V(\widetilde{P})$.
\end{Lemma}

{\it Proof} is evident.

\begin{Corollary}\label{c3}
Let $G$ be a graph, $\varphi\in\lambda(G)$, $\{x,y\}\subseteq
V''(G^{(\varphi,int)})$, $(x,y)\in E(G)$. Then there exists a unique
$\widetilde{P}\in\tau(G,\varphi)$ satisfying the condition
$\{x,y\}\subseteq V(\widetilde{P})$.
\end{Corollary}

\section{Main Result}

\begin{Theorem}\cite{Luhansk}
For any graph $G$ and arbitrary $\varphi\in\lambda(G)$,
$G^{(\varphi,int)}$ is a forest, each connected component $H$ of
which satisfies one of the following two conditions: 1) $H\cong
K_1$, and the only vertex of the graph $H$ may or may not belong to
the set $V'(G)$, 2) $H$ is a galaxy satisfying one of the following
three conditions: a) $V'(H)\subseteq V'(G)$, b) exactly one vertex
of the set $V'(H)$, which is a peripheral vertex of $H$, doesn't
belong to the set $V'(G)$, c) exactly two vertices of the set
$V'(H)$, with $diam(H)$ as the distance between them, don't belong
to the set $V'(G)$.
\end{Theorem}

{\it Proof.} Choose an arbitrary $\varphi\in\lambda(G)$. Let us
consider an arbitrary connected component $H$ of the graph
$G^{(\varphi,int)}$. By corollary \ref{c2}, $H$ is a tree.

{\it Case 1.} $|V(H)|=1$. In this case there is nothing to prove.

{\it Case 2.} $|V(H)|=2$. Clearly, $H\cong K_2$, and the proposition
is evident.

{\it Case 3.} $|V(H)|\geq3$. Clearly, $|V''(H)|\geq1$.

{\it Case 3.1.} $|V''(H)|=1$. In this case $H\cong K_{\Delta(H),1}$,
$|V'(H)|=\Delta(H)=|V(H)|-1\geq2$, $diam(H)=2$. Clearly, $H$ is a
galaxy, and all vertices of $V'(H)$ are peripheral vertices of the
graph $H$. Without loss of generality we can assume that
$V''(H)=\{u_0\}$, and, moreover, that the vertices $u'\in V'(H)$ and
$u''\in V'(H)$ satisfy the conditions
$l(\varphi[E(H)])=\varphi((u_0,u'))$,
$L(\varphi[E(H)])=\varphi((u_0,u''))$.

Now consider an arbitrary vertex $z\in V'(H)$, which satisfies the
condition $l(\varphi[E(H)])<\varphi((u_0,z))<L(\varphi[E(H)])$.

Let us show that $z\in V'(G)$. Assume the contrary: $z\in V''(G)$.
From here we obtain the inequality $|\varphi[J_G(u_0)]\cap
\varphi[J_G(z)]|\geq2$, which contradicts lemma \ref{l1}.

Consequently, $V'(H)\cap V''(G)\subseteq\{u',u''\}$, and, therefore,
$0\leq|V'(H)\cap V''(G)|\leq2$. It completes the proof of case 3.1.

{\it Case 3.2.} $|V''(H)|\geq2$. Clearly, in this case there exist
vertices $x\in V''(H)$ and $y\in V''(H)$ satisfying the condition
$(x,y)\in E(H)$. By corollary \ref{c3}, there exists a unique
$P_0\in\tau(G,\varphi)$ with $\{x,y\}\subseteq V(P_0)$. Suppose that
$w'$ and $w''$ are endpoints of $P_0$. It is not difficult to see
that $d_{P_0}(w',w'')\geq3$ and $|I_G(w')\cap
U(G,\varphi)|=|I_G(w'')\cap U(G,\varphi)|=1$.

Let us show that
$$
\Big(\bigcup_{x\in V''(P_0)}I_H(x)\Big)\backslash V(P_0)\subseteq
V'(G).
$$

If $(\bigcup_{x\in V''(P_0)}I_H(x))\backslash V(P_0)=\emptyset$, the
required relation is evident. Now assume that $(\bigcup_{x\in
V''(P_0)}I_H(x))\backslash V(P_0)\neq\emptyset$.

Choose an arbitrary vertex $z\in(\bigcup_{x\in
V''(P_0)}I_H(x))\backslash V(P_0)$. Let us show that $z\in V'(G)$.
Assume the contrary: $z\in V''(G)$.

Consider the vertex $z_0\in V''(P_0)$ which is adjacent to $z$. From
the properties of $P_0$ it follows that
$l(\varphi[J_H(z_0)])<\varphi((z_0,z))<L(\varphi[J_H(z_0)])$. Since
$\{z_0,z\}\subseteq U(G,\varphi)$ we obtain that
$|\varphi[J_G(z_0)]\cap \varphi[J_G(z)]|\geq2$. It contradicts lemma
\ref{l1}.

Thus, indeed, $(\bigcup_{x\in V''(P_0)}I_H(x))\backslash
V(P_0)\subseteq V'(G)$. It implies that $V'(H)\cap
V''(G)\subseteq\{w',w''\}$, $0\leq|V'(H)\cap V''(G)|\leq2$, and,
that $P_0$ is the unique path in the graph $H$ between its vertices
$w'$ and $w''$. Now it is easy to see that
$diam(H)=diam(P_0)=d_{P_0}(w',w'')=d_H(w',w'')$. It completes the
proof of case 3.2.

{\it Theorem is proved.}

\begin{Corollary}\label{c4}\cite{Luhansk}
If $G$ is a graph with $V'(G)=\emptyset$, and
$\varphi\in\lambda(G)$, then an arbitrary connected component of the
forest $G^{(\varphi,int)}$ is a simple path.
\end{Corollary}

\begin{Corollary}\label{c5}\cite{Luhansk}
A labeling, which provides every vertex of a graph $G$ with an
interval spectrum, exists iff $G$ is a galaxy.
\end{Corollary}

\begin{Corollary}\label{c6}
If $n\in \mathbb{N}$, and $\varphi\in\lambda(K_n)$, then the forest
$K_n^{(\varphi,int)}$ is a tree which is isomorphic either to $K_1$,
or to $K_2$.
\end{Corollary}

{\bf Received: Month xx, 20xx}

\end{document}